# Total dose radiation and annealing responses of the back transistor of Silicon-On-Insulator pMOSFETs

ZHAO Xing(赵 星)[1;1)]　ZHENG Zhong-Shan(郑中山)[1;2)]　LI Bin-Hong(李彬鸿)[1]

GAO Jian-Tou(高见头)[1)]　YU Fang(于 芳)[1]

[1] Institute of Microelectronics, Chinese Academy of Sciences, Beijing 100029, China

**Abstract:** The total dose radiation and annealing responses of the back transistor of Silicon-On-Insulator (SOI) pMOSFETs have been studied by comparing with those of the back transistor of SOI nMOSFETs fabricated on the same wafer. The transistors were irradiated by $^{60}$Co γ-rays with various doses, and the front transistors are biased in a Float-State and Off-State, respectively, during irradiation. The total dose radiation responses of the back transistors are characterized by their threshold voltage shifts. The results show that the total dose radiation response of the back transistor of SOI pMOSFETs, similar to that of SOI nMOSFETs, depends greatly on their bias conditions during irradiation. However, with the Float-State bias, rather than the Off-State bias, the back transistors of SOI pMOSFETs reveal a much higher sensitivity to total dose radiation, which is contrary to those of SOI nMOSFETs. In addition, it is also found that the total dose radiation effect of the back transistor of SOI pMOSFETs irradiated with Off-State bias, as well as that of the SOI nMOSFETs, exhibits an increasing trend as the channel length decreases. The annealing response of the back transistors after irradiation at room temperature without bias, characterized by their threshold voltage shifts, indicates that there is a relatively complex annealing mechanism associated with channel length and type, bias condition during irradiation. In particular, for all the transistors irradiated with Off-State bias, their back transistors show an abnormal annealing effect during early annealing. All the results have been discussed and analyzed in detail by the aid of simulation.

**Key words**：SOI pMOSFET, back transistor, total dose radiation, annealing

**PACS**：61.80.Az, 61.80.Ed, 73.40.Qv

1) E-mail: zhaoxing1@ime.ac.cn

2) E-mail: zhengzhongshan@ime.ac.cn

1. **Introduction**

The total dose radiation responses on Silicon-On-Insulator (SOI) devices and circuits have always been one of the focuses of SOI technology.[1-7] With the advantages in increasing circuit speed, lowering power consumption and improving circuit integration density, SOI technology has great potential to promote continuously the development of integrated circuit (IC) technology and becomes one of mainstream IC technologies in the future. However, for space applications, SOI devices and circuits are more sensitive to total dose radiation due to the buried oxide (BOX) layer in SOI materials. Moreover, the total dose radiation responses of SOI devices is closely related to bias conditions during irradiation, device sizes, device structures, and manufacturing processes, etc, making the total dose radiation response and hardening complex.[8-10] As the device dimension continuously decreases and the gate oxide thickness is scaled down, the threshold voltage shift induced by the gate oxide radiation damage has already become negligible in deep submicron IC technology.[11] Therefore, the BOX radiation damage dominates in the total dose radiation response of SOI devices and ICs, which can cause not only the threshold voltage shift of the parasitic back transistor of SOI MOSFETs, but also that of the front transistor by electrical coupling between the front and back transistor for fully-depleted (FD) SOI MOSFETs.[12] Although a lot of work on the total dose radiation effect of SOI devices has been done,[13-15] little effort has been made for SOI pMOSFETs, for the reason that the increasing leakage current of SOI circuits in radiation environments is mainly due to the radiation-damaged n-channel SOI devices. However, with the development of SOI technology, especially, FD SOI technology as a very promising candidate overcoming the device scaling challenges, the total dose radiation effect of SOI pMOSFETs will be of concern. Even though the BOX radiation damage in SOI pMOSFETs does not usually bring an increase in leakage current, its influence on the front device of FD p-channel SOI devices will become crucial, because of the enhanced electrical coupling between the front and back device, with the body thickness decreasing, resulting in some other non-negligible effects on related circuit performances besides the increase in static power consumption. In view of this, the work of this paper focuses on the total dose radiation response of the back transistors of SOI pMOSFETs, so as to obtain some useful results that contribute to a comprehensive understanding of the radiation effects of SOI technology, especially those of FD SOI technology. To avoid the effect of the front transistor on the back one due to the electrical coupling, SOI pMOSFETs used for this work were fabricated with partially-depleted (PD) SOI technology. Also, SOI nMOSFETs were fabricated on the same wafer for comparison. In addition, the annealing response of the back transistors at room temperature has been observed after irradiation, without bias during annealing.

## 2. Description of Devices and Experiments

The SOI transistors in this work, for both the pMOSFETs and the nMOSFETs, with gate lengths being respectively 8.0μm, 1.6μm and 0.8μm, were fabricated using a 0.8μm standard PD SOI CMOS technology on the separation by implanted oxygen (SIMOX) SOI wafer with the top silicon and the buried oxide layer being 235nm and 375nm, respectively. In addition, all the fabricated transistors have a gate oxide thickness of 12.5nm, a gate width $W$ of 8μm, and a body contact for body biasing. The great difference between the gate lengths can ensure that there are observable differences in the total dose radiation response due to various gate lengths. So, the influence of the gate length $L$ on the total dose radiation response can also be observed in this work, besides irradiation dose and bias.

Both the pMOSFETs and the nMOSFETs were divided into two groups, respectively, for two different irradiation biases: Off-State and Float-State, as specified and summarized in Table 1, with the symbols $V_S$, $V_G$, $V_D$, $V_B$, and $V_{SUB}$ denoting source, gate, drain, body, and substrate bias, respectively, and "×" denoting no bias. In particular, Float-State used as a control, which actually means that no voltage is applied to each of the terminals of the transistors, can give a bias reference, showing clearly the irradiation bias effect on the total dose radiation response.

Table 1. Bias configurations used in this work during irradiation

| SOI Transistors and Bias Configurations | $V_S$ / V | $V_G$ / V | $V_D$ / V | $V_B$ / V | $V_{SUB}$ / V |
| --- | --- | --- | --- | --- | --- |
| pMOSFET/Off-State | 5 | 5 | 0 | 5 | 0 |
| pMOSFET/Float-State | × | × | × | × | × |
| nMOSFET/Off-State | 0 | 0 | 5 | 0 | 0 |
| nMOSFET/Float-State | × | × | × | × | × |

First, the transfer characteristic of the back transistor of the SOI MOSFETs was measured at room temperature with a Keithley 4200-SCS semiconductor parameter analyzer before irradiation. Then, the transistors were irradiated with Off-State and Float-State bias, respectively, using [60]Co gamma rays at a dose rate of 50 rad(Si)/s with six doses: 20, 50, 100, 200, 300, and 500 krad(Si). After each dose irradiation, the transfer characteristic of the back transistors was immediately measured again at room temperature with the same parameter analyzer. The total dose radiation response of the back transistors is characterized by their threshold voltage shifts due to irradiation, extracted from the measured transfer characteristics.

After 500 krad(Si) irradiation, the annealing response of the back transistors of the SOI pMOSFETs at room temperature without bias was observed and characterized by their threshold voltage shifts as a function of time during annealing.

## 3. Results of Experiments

For the two different irradiation biases, i.e. Float-State and Off-State, Figure 1 and Figure 2 illustrate the typical transfer characteristic curves of the back transistor of the SOI pMOSFETs before and after irradiation, measured at a source-drain voltage $V_{SD}$ of 0.1V, with $I_D$ denoting drain current and $V_{G(B-G)}$ denoting back gate (B-G) bias. According to Figures 1 and 2, it is obvious that, although the threshold voltage shift of the back transistors increases with increasing irradiation dose for both Float-State and Off-State bias, the back transistor of the SOI pMOSFETs irradiated with Float-State bias has a much greater shift for each dose, showing a higher sensitivity to total dose radiation. Further, for the SOI pMOSFETs with the three different channels, Figure 3 shows the threshold voltage shift $\Delta V_{th\,(B-G)}$ of their back transistors due to irradiation with the two different biases as a function of radiation dose $D$, obtained from the measured transfer characteristics. It is clear that, for the SOI pMOSFETs irradiated with Float-State bias, all the back transistors reveal much bigger shifts, compared with irradiation with Off-State bias, in spite of the differences due to channel length for the same bias. Additionally, for comparison, Figure 4 gives the radiation-induced $\Delta V_{th\,(B-G)}$ of the back transistor of the SOI nMOSFETs biased with Off-State and Float-State, respectively, during irradiation, showing that the greater $\Delta V_{th\,(B-G)}$ occurs under Off-State bias rather than Float-State one, contrary to the SOI pMOSFETs.

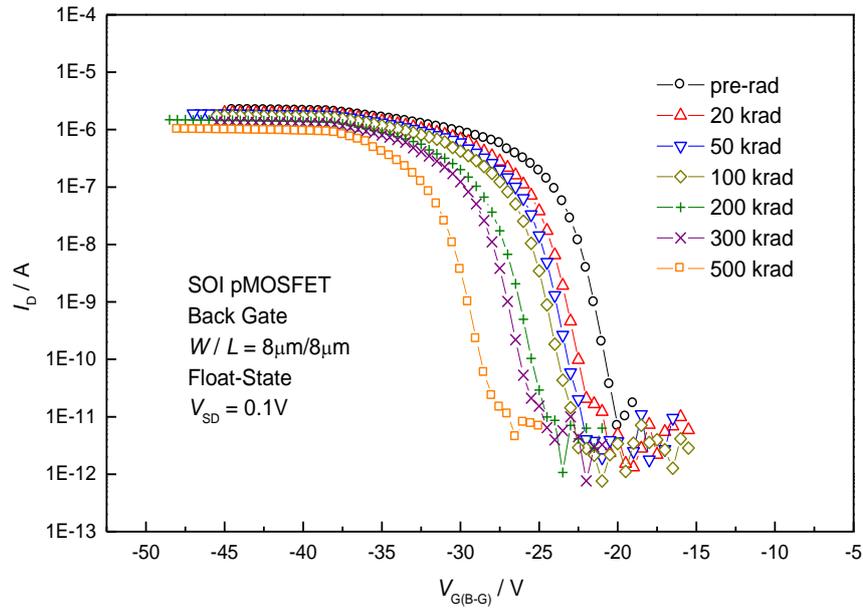

Fig. 1. The transfer characteristic curves of the back transistor of SOI pMOSFETs (*W/L*=8μm/8μm) biased with Float-State during irradiation, before and after irradiation.

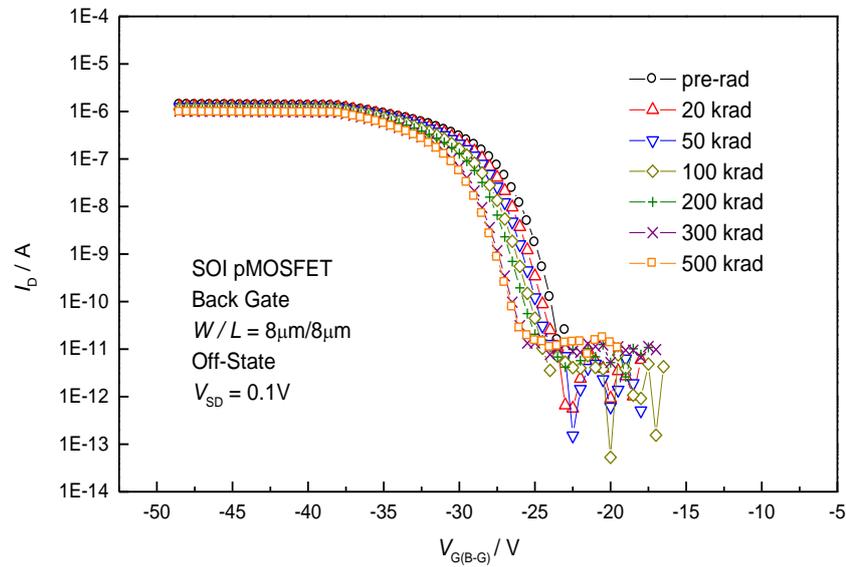

Fig. 2. The transfer characteristic curves of the back transistor of SOI pMOSFETs (*W/L*=8μm/8μm) biased with Off-State during irradiation, before and after irradiation.

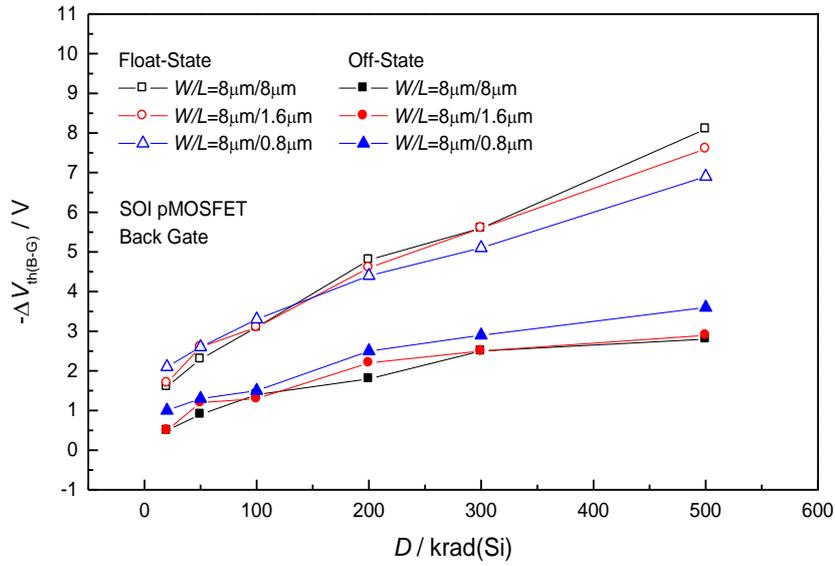

Fig. 3. The radiation-induced threshold voltage shift $\Delta V_{th\,(B\text{-}G)}$ of the back transistor of the SOI pMOSFETs with the three different channels biased with Float-State and Off-State, respectively, during irradiation, as a function of radiation dose $D$.

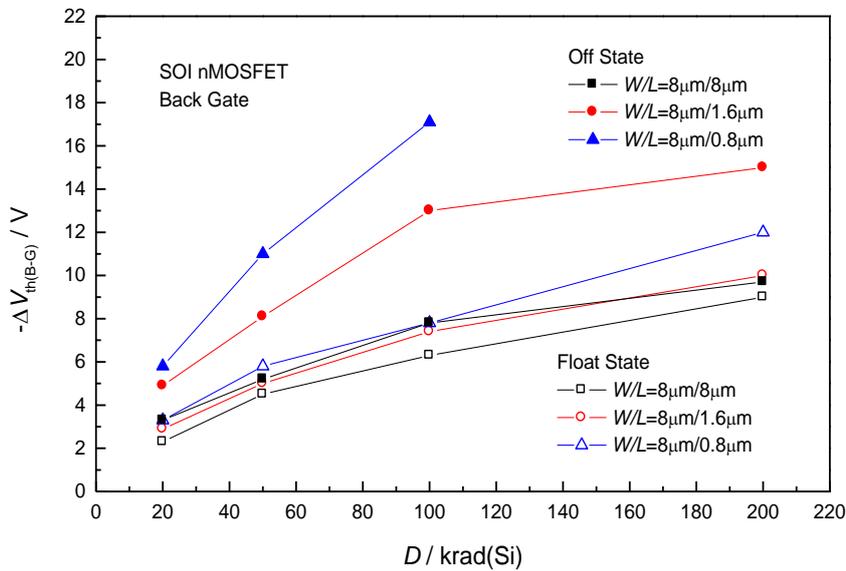

Fig. 4. The radiation-induced threshold voltage shift $\Delta V_{th\,(B\text{-}G)}$ of the back transistor of the SOI nMOSFETs with the three different channels biased with Float-State and Off-State, respectively, during irradiation, as a function of radiation dose $D$.

Figures 5 and 6 show the typical transfer characteristic shift of the back transistor of the SOI pMOSFETs due to annealing at room temperature without bias after 500 krad(Si) irradiation. For irradiation with Float-State bias, the back transistor in Figure 5 displays a normal annealing response, i.e. a positive transfer characteristic or threshold voltage shift with increasing annealing time. However, with Off-State bias, the back transistor in Figure 6 has a significant negative transfer characteristic shift in the early stage of annealing, showing an anomalous annealing response related closely to the bias configuration during irradiation. As a function of annealing time $t$, Figure 7 shows the threshold voltage $V_{th}$ curves of the back transistors of the SOI pMOSFETs, and Figure 8 those of the SOI nMOSFETs for comparison. It is obvious that, for irradiation with Off-State bias, both p- and n-channel back transistors all exhibit negative threshold voltage shifts in varying degrees during early annealing. On the other hand, with Float-State bias, the back transistors basically have a normal annealing response, but a $V_{th}$ rebound in Figure 7 during latter annealing.

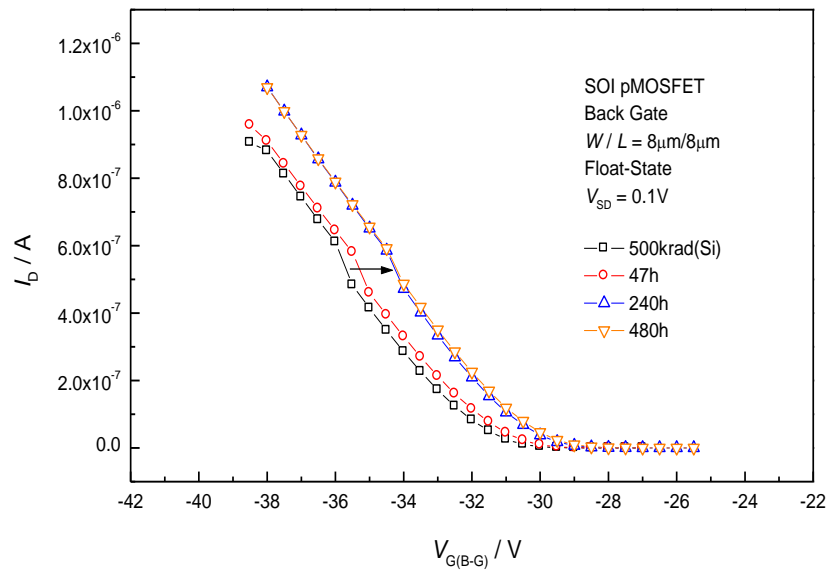

Fig. 5. The annealing effect of the back transistor of the SOI pMOSFET ($W/L$=8μm/8μm) irradiated with Float-State bias, at room temperature without bias.

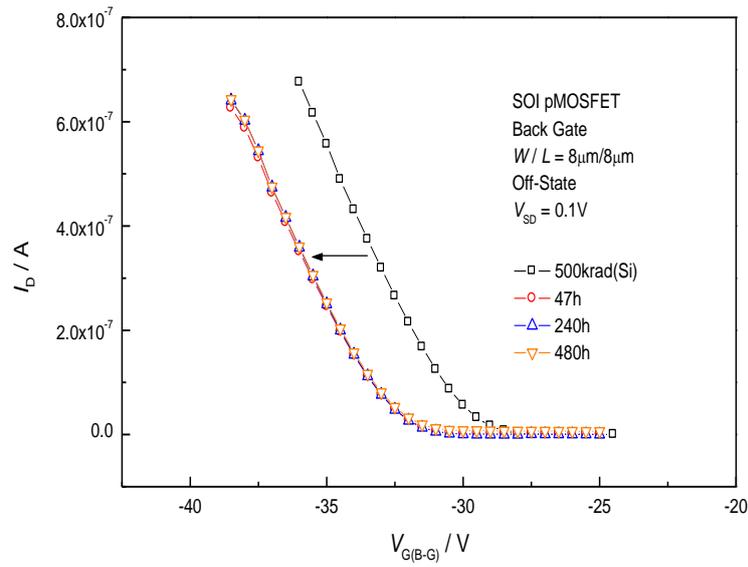

Fig. 6. The annealing effect of the back transistor of the SOI pMOSFET (*W/L*=8μm/8μm) irradiated with Off-State bias, at room temperature without bias.

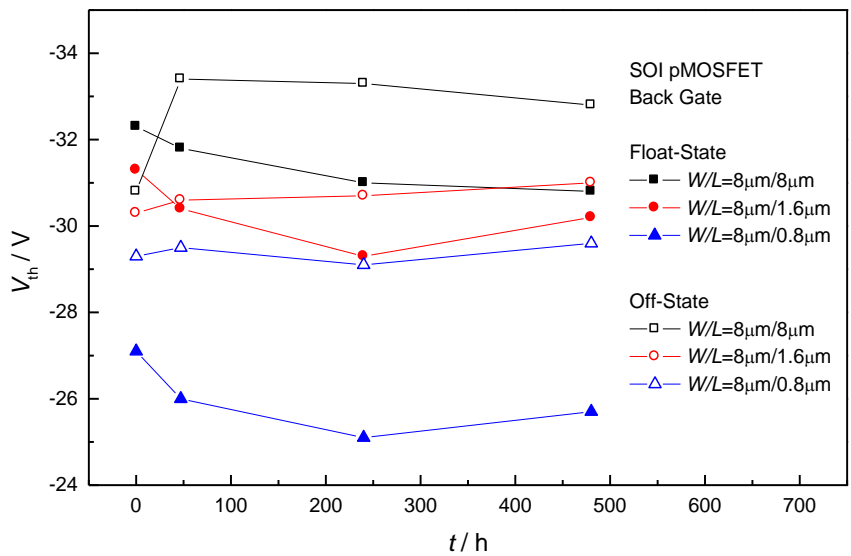

Fig. 7. The threshold voltage $V_{th}$ of the back transistor of the SOI pMOSFETs irradiated with Float-State and Off-State, respectively, as a function of annealing time *t*.

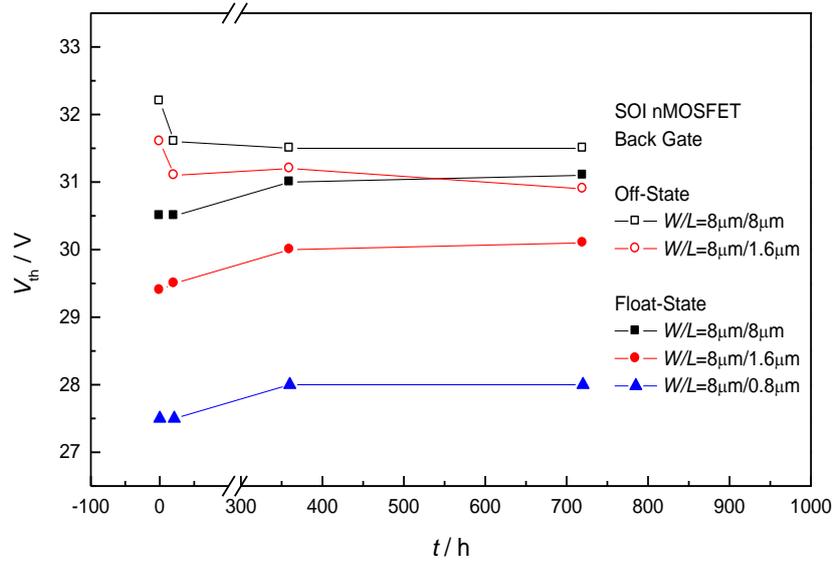

Fig. 8. The threshold voltage $V_{th}$ of the back transistor of the SOI nMOSFETs irradiated with Float-State and Off-State, respectively, as a function of annealing time $t$.

## 4. Discussion

As shown in Figure 3, for irradiation with Float-State bias, the threshold voltage shift of the back transistor of the SOI pMOSFETs is much greater, compared with Off-State bias. The great dependence of the radiation response on bias configuration during irradiation can be explained by Figure 9, which illustrates a simulated two-dimensional potential and electric field distribution in the BOX for a SOI pMOSFETs with Off-State bias. From Figure 9, it is clear that, under the back channel, there is an electric field pointing towards the BOX-substrate interface, which will push holes generated in the BOX during irradiation to move to the BOX bottom. As a result, the distribution center of radiation-induced trapped holes in the BOX will tend to the BOX bottom away from the back channel, reducing greatly the influence of the trapped holes on the back channel, and showing a much smaller threshold voltage shift of the back transistors. As an example, Figures 10 and 11 show schematically the simulated hole concentration distributions 10nm below the body-BOX interface (i.e. the top BOX interface) and 10nm above the BOX-substrate interface (i.e. the bottom BOX interface) during irradiation with Off-State and Float-State bias, respectively, for the same SOI pMOSFET as in Figure 9, illustrating how the electric field in the BOX or bias configuration affects the distribution of trapped holes in the BOX layer.

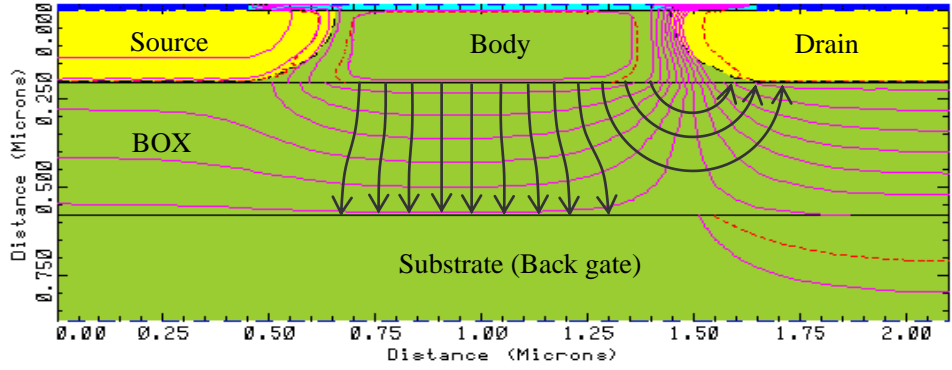

Fig. 9. The Simulated potential and electric field distribution in BOX under the back channel of a SOI pMOSFET with $L = 0.8\mu m$ biased with Off-State condition.

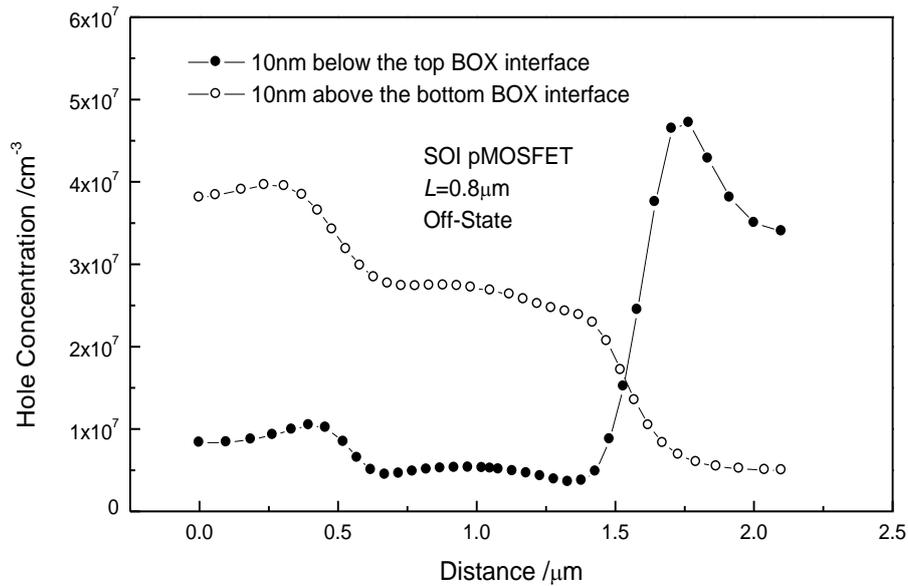

Fig. 10. The simulated hole concentration distributions in BOX 10nm below the top BOX interface and 10nm above the bottom BOX interface for a SOI pMOSFET with $L = 0.8\mu m$ during irradiation with Off-State bias.

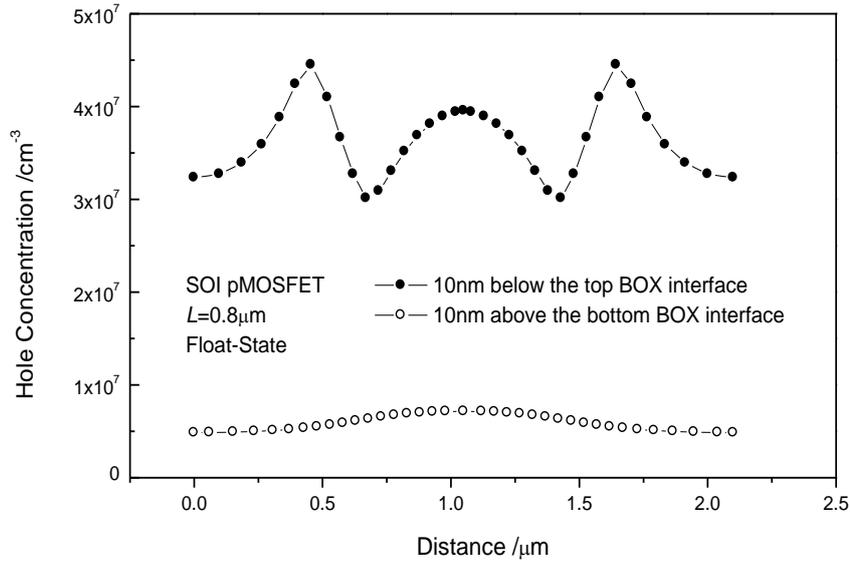

Fig. 11. The simulated hole concentration distributions in BOX 10nm below the top BOX interface and 10nm above the bottom BOX interface for a SOI pMOSFET with $L = 0.8\mu m$ during irradiation with Float-State bias.

On the other hand, from Figures 1 and 2, there are no observable radiation-induced subthreshold slope distortions, illustrating that the effect of the body-BOX interface traps generated during irradiation on the threshold voltage of the back transistors is negligible, and the back transistor threshold voltage shift is mainly caused by the trapped charges in the BOX due to irradiation.

In addition, Figure 3 also shows that, for irradiation with Float-State bias, the $0.8\mu m$ back transistor (i.e. the back transistor corresponding to a front channel length of $L = 0.8\mu m$ ) has a bigger threshold voltage shift in the early stages of irradiation and a smaller one in the later stages, compared with the other two back transistors. This can be attributed to the change of buildup of trapped holes under the source, drain, and back channel in the BOX with increasing irradiation dose. With $L=0.8\mu m$, i.e. the shortest back channel, the trapped holes under the source and drain will have the strongest effect on the back channel average surface potential due to the shortest average distance from the source or drain to the back channel. So, during early irradiation, when the trapped holes under the source and drain increase rapidly and dominate the back transistor threshold voltage shift, the back transistor with the shortest channel length reveals the highest radiation sensitivity. The simulated hole concentration distribution curves in Figure 11 support this analysis, which shows a higher hole concentration under the source and drain near the back channel during irradiation. However, when the neutral BOX traps decrease gradually due to hole trapping, and the increase of trapped holes under the source and drain with irradiation is suppressed to a greater extent due to early irradiation, the trapped holes under the back channel, instead of those under the source and drain, will become a main factor caused a further back transistor threshold voltage shift with increasing irradiation dose. Since there are fewer trapped holes under a shorter back channel, the 0.8μm back transistor shows the smallest threshold voltage shift due to its shortest channel length during late irradiation.

From Figure 3, it can also be seen that, when the irradiation bias is the Off-State configuration, the 0.8μm back transistor always has the biggest threshold voltage shift among the three back transistors throughout the irradiation. Similarly, this can also be explained by hole trapping related to hole concentrations in the BOX during irradiation. As seen from Figure 10, under the back channel, there are lower hole concentrations near the top BOX interface; under the drain, however, much higher ones. Thus, the radiation induced hole-trapped density under and near the drain will be much higher than that under and near the back channel. For a short back channel, just analyzed previously, its threshold voltage or average surface potential is more sensitive to those trapped holes under the drain, compared to a long one. Meanwhile, the increase of trapped holes under and near the back channel, which plays a more important role in the threshold voltage shift of long channel back transistors than that of short channel ones, is suppressed by the electric field in the BOX. So, the back transistor with the shortest channel 0.8μm always displays the biggest threshold voltage shift for each irradiation dose.

However, from Figure 4, the n-channel back transistors reveal essentially their higher radiation sensitivities for the Off-State case, compared with the Float-State one, showing that, for the p- and n-channel back transistors, there is a reverse radiation sensitivity dependence on the irradiation bias. It is also found in Figure 4 that there is a great difference between the radiation responses of the back transistors with different channel lengths for irradiation with Off-State bias, and a shorter back channel corresponds to a greater threshold voltage shift as expected. To have an insight into these, the electric potential and electric field distributions in the BOX under the back channel of a 0.8μm channel SOI nMOSFET biased with Off-State are simulated and shown in Figure 12. It is obvious that, near the top BOX interface, there is an electric field pointing towards the back channel in the BOX, which is very different from the SOI pMOSFET case in Figure 9, leading to a hole accumulation near the top BOX interface under the back channel during irradiation, as shown in Figure 13. As a result, the probability that the BOX hole traps near the back channel capture holes is enhanced, resulting in a rapid increase of trapped holes near the back channel. From figure 12, it is also clear that the drain voltage of the Off-State configuration strengthens the electric field under the back channel. So, for the Off-State bias, the back transistors will have bigger threshold voltage shifts due to more trapped holes under the back channel, compared with the Float-State case without bias, just as in Figure 4.

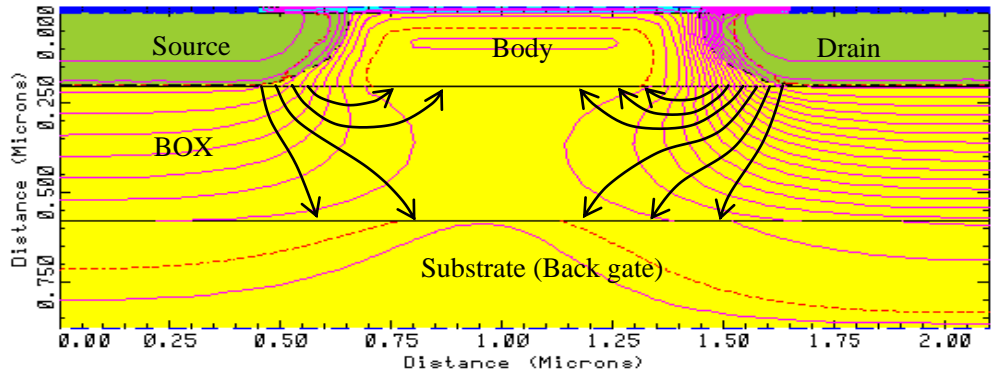

Fig. 12. Two-dimensional potential and electric field distribution in the BOX under the back channel of a SOI nMOSFET with 0.8μm channel under the Off-State bias condition.

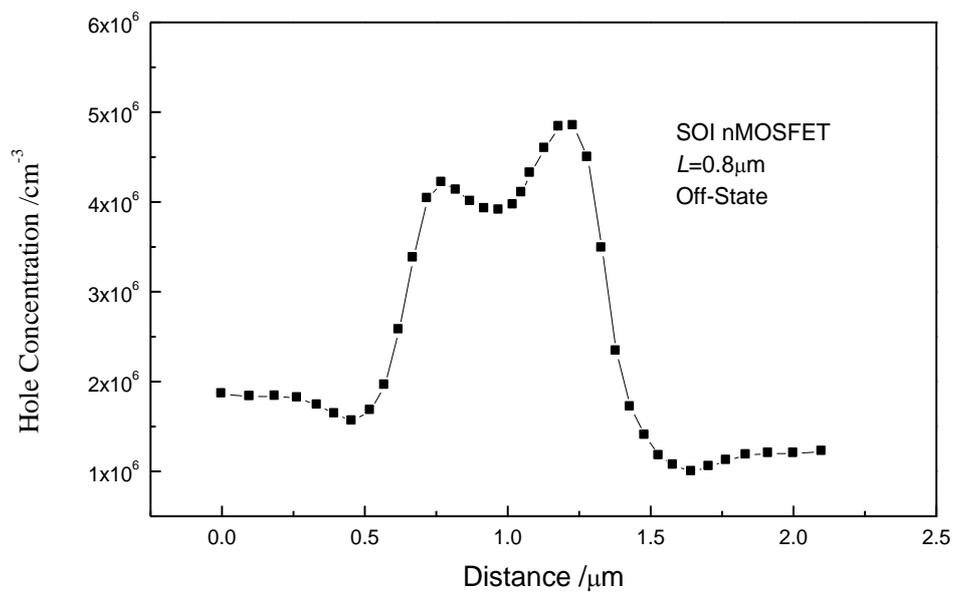

Fig. 13. The simulated BOX hole concentration distribution 10nm under the top BOX interface for a 0.8μm channel SOI nMOSFET under the Off-State bias condition during irradiation.

On the other hand, according to Figure 12, a longer back channel will be helpful in reducing the effect of electric field on the BOX region near the back channel, especially near the middle region of the back channel. Thus, the 8μm back transistor exhibits the smallest threshold voltage shift, as showed in Figure 4. Furthermore, it is evident that the trapped holes which are exactly under the back channel have a much greater effect on the back channel threshold voltage than those under the drain. Therefore, for irradiation with Off-State bias, the different n-channel back transistors display their great radiation response differences due to the great difference between their channel lengths.

Typically, the radiation damage in MOSFETs can be partly removed by annealing with a much higher temperature than room temperature, and the threshold voltage shows a positive shift due to annealing as a result. However, from Figures 7 and 8, after irradiation with Off-State bias, the back transistors all reveal an unusual annealing effect of radiation damage, i.e. the negative shift of the threshold voltages, in the early annealing stage at room temperature without bias. This seems to indicate the production of an additional damage in the BOX, similar to the enhancement of radiation damage, due to annealing. Subsequently, the back transistor threshold voltage shifts from such an annealing show that the annealing effect is related to not only the bias during irradiation, but also the channel length and type.

Since the annealing of radiation damage is essentially a process of the reduction or recombination of radiation-induced trapped charges, it is impossible that the trapped charges in the BOX significantly increase, just as for irradiation, because of the annealing at room temperature with no bias, and bring a quite apparent negative shift of the back transistor threshold voltage. Therefore, a probable reason for the unusual annealing effect described above is because of the non-uniform distribution of the trapped holes in the BOX due to irradiation with Off-State bias, which can result in the diffusion or redistribution of the trapped holes in the BOX during annealing, and contribute to the back transistor threshold voltage shift. For example, for the SOI pMOSFETs irradiated with Off-State bias, when the trapped holes under the drain, which have a higher concentration from Figure 10, diffuse or spread to the region under the back channel due to the removal of the external Off-State bias and the change of the internal electric field in the BOX, and this diffusion effect on the back transistor threshold voltage is dominant over that of the recombination of the trapped charges during annealing, which is probable at room temperature, a negative shift of the back transistor threshold voltage will occur due to the annealing. Similarly, the other negative shift cases, including the rebound of the back transistor threshold voltage in the latter stage of the annealing, can be explained by the redistribution of the trapped charges in the BOX. To obtain the more detailed mechanisms for them, a further study is needed.

## 5. Conclusion

In conclusion, the total dose radiation response of the back transistor of SOI pMOSFETs depends more greatly on bias voltages during irradiation, compared with channel lengths. In particular, for the SOI pMOSFETs irradiated with Float-State bias, their back transistors show much higher radiation sensitivity than for the Off-State case, which is unexpected to some extent and contrary to the results on the irradiated SOI nMOSFETs. After irradiation, the threshold voltage shift of the back transistor of the SOI MOSFETs during annealing without bias indicates a relatively complex room-temperature annealing response. In particular, the occurrence of an unusual negative threshold voltage shift due to room-temperature annealing, which is attributed to the diffusion of trapped charges in the BOX during annealing in this paper, reflects some special annealing mechanism of radiation damage, which is possible to be obscured during higher-temperature annealing, and a further study is needed to gain an insight into this.